# A comparative study of resists and lithographic tools using the Lumped Parameter Model

Roberto Fallica[a)], Robert Kirchner, Yasin Ekinci

Paul Scherrer Institut, 5232 Villigen PSI, Switzerland

Dominique Mailly

Laboratoire de photonique et de nanostructures, CNRS, Route de Nozay, 91460 Marcoussis, France

[a)] Electronic mail: roberto.fallica@psi.ch

A comparison of the performance of high resolution lithographic tools is presented here. We use extreme ultraviolet interference lithography, electron beam lithography, and He ion beam lithography tools on two different resists that are processed under the same conditions. The dose-to-clear and the lithographic contrast are determined experimentally and are used to compare the relative efficiency of each tool. The results are compared to previous studies and interpreted in the light of each tool-specific secondary electron yield. In addition, the patterning performance is studied by exposing dense line/spaces patterns and the relation between critical dimension and exposure dose is discussed. Finally, the Lumped Parameter Model is employed in order to quantitatively estimate the critical dimension of line/spaces, using each tool specific aerial image. Our implementation is then validated by fitting the model to the experimental data from interference lithography exposures, and extracting the resist contrast.

# I. INTRODUCTION

Over the last four decades, several novel lithographic techniques such as electron beam lithography[1,2], nanoimprint lithography[3], projection and focused ion beam lithography[4,5] have been developed and have become widely employed in research labs and manufacturing fabs. Among these equipment, some are economically viable and technologically mature for high-volume manufacturing as, for example, photolithography and nanoimprint lithography. Conventional optical lithography has been the workhorse of high-volume manufacturing and has steadily evolved in terms of wavelength and new concepts, e.g. ArF immersion[6] and extreme ultraviolet (EUV) lithography[7]. Other techniques, such as those based on electron beam and on ion beam direct writing, are not yet suitable for mass production but are still of paramount importance for specific applications such as fabrication of masks for optical lithography, specific ultrahigh resolution applications, and for academic research and ad hoc nanofabrication. The field of lithography in a general sense is in continuous evolution on several fronts. The recent introduction of gaseous Ne and He ion sources has been a significant breakthrough[8] owing to the possibility of achieving extremely small virtual source size ($\approx$ 0.25 nm) while, at the same time, providing relatively high beam current. These two features make it desirable not only for high resolution microscopy, but especially for lithography, patterning and nanofabrication, as plenty of studies have shown.[9,10,11,12]

In any lithography process, the resist chemistry plays a paramount role because the lithographic performance is coupled to the imaging performance of the material and the lithographic equipment. During exposure to photons, electrons or ions, resist molecules undergo chemical bond formation, in a negative tone resist, or scission, in a positive tone resist. The core of any lithographic process is the way an aerial image (i.e. the spatial distribution of energy) produces the resist image (i.e. the final pattern or feature). Finding the relation between the former and the latter and its accurate modeling is of substantial interest for estimation and optimization of any lithographic process. Nevertheless, this undertaking is far from being straightforward and several analytical models have been developed and are available for simulating the exposure by electron beam[13], by optical lithography[14,15], and by X-ray lithography[16]. There are many other so called "full resist models", which provide higher accuracy and versatility, at the expense of computational burden: these are being commercialized in software suites (Dr. LiTHO, PROLITH, TRAVIT, Sentaurus). These comprehensive models can estimate several process parameters, such as the resist dissociation rate, the deblocking fraction, the solubility change, etc. The wealth of lithographic models and software testifies to the technological importance of lithographic modeling.

Understanding of the lithographic process and the accurate modeling thereof become more and more important towards achieving cutting-edge resolution in the fabrication of nanoscale devices. This undertaking is a great challenge not only for large-scale industrial production, but also for academic purposes and prototyping. The state-of-the-art equipment have strengths and weaknesses on their own. The electron beam lithography (EBL) for instance is a well-established technique but is relatively slow and suffers from large proximity effect. This effect is, in general, taken into account by tuning the exposure dose according to the geometry and most EBL design software provide such compensation. The He ion beam lithography (HIBL) is a technique that holds promise for higher resolution and better lithographic control than EBL, owing to the reduced proximity effect. Optical lithography brings the advantage of fast patterning and high density but its resolution is limited by the wavelength used.

For these reasons, in this work we compare and discuss the performance of three different tools: EUV interference lithography (EUV-IL), EBL, and HIBL in patterning a periodic layout of densely packed lines and spaces (l/s). L/s are ubiquitously used in integrated circuit architecture such as crossbar memory devices, metal lines, programmable logic arrays, word & bit lines; and they will likely be also used in future 3-D devices.[17] Moreover, dense l/s represent the ultimate resolution testing condition for resist and tools where the proximity effect is accounted for. In the present work, we focus on the amount of energy (or dose) required to print a resist feature of given size (critical dimension, CD), also known as the CD vs. dose function, which is a relevant figure of merit for lithographers. Although the exposure mechanism is radically different in these tools, the exposure chemistry is always triggered by the (primary or secondary) electrons in all the three tools. We discuss how these differences affect the final result also in the light of existing studies on the exposure dosage of photoresists by EBL and EUV. Finally, an implementation of the lumped parameter model (LPM) in the case of a periodic l/s pattern is presented. This model is employed to present a quantitative comparison of the effect of the aerial image of different tools on the CD vs. dose relationship. Our implementation is then validated on experimental data from optical lithography by using a nonlinear least square regression.

## II. EXPERIMENTAL

Here, we describe the characteristics of the three lithographic tools and resist materials used in this work. The conventional geometry adopted here is that the l/s patterns are printed on the resist along one direction parallel to the surface of the sample. It is assumed $z$ along the normal to the surface of the sample. The aerial image intensity, *I* generated by each exposure tool is thus symmetrical along the direction of lines and it is fully described as a function of *x* only. The CD of lines was experimentally measured from by top-down SEM imaging of the patterned resist l/s and by quantitative metrology image analysis using a commercial software suite (SuMMIT, Lithometrix).

## A. *Extreme Ultraviolet-Interference Lithography*

The extreme ultraviolet interference lithography tool (EUV-IL) at the XIL-II beamline uses light at 13.5 nm wavelength from the Swiss Light Source. Masks featuring transmission diffraction gratings produce two-beam interference patterns on the surface of the sample to form dense l/s, as detailed elsewhere.[18] The main advantages of this technique are the high resolution (down to 6 nm half-pitch) and the fast speed to pattern large areas.[19] The shape of the aerial image is dictated by the constructive and destructive interference of the two diffracted beams and it is given by:

$$I_{EUV-IL}(x) = \frac{1}{2}\cos\left(2\pi\frac{x}{p}\right) + \frac{1}{2} = cos^2(\pi\frac{x}{p}) \qquad (1)$$

where *p* is the pitch of the l/s array.

For the two-beam interference l/s patterning, pairs of gratings of different pitches were fabricated on SiN membranes. For producing the contrast curves, an aperture of 0.5×0.5 mm$^2$ was used to expose an array of increasing doses.

## B. *Electron beam lithography*

An electron beam lithography tool (Vistec EBPG 5000) with 100 keV acceleration voltage, beam current of 500 pA at aperture 400 μm was used for the present study. The direct write with a raster-scan focused beam has a relatively low speed and can print only about a few μm$^2$ per second, depending on a variety of settings. For the sake of this comparison, we are interested in the effect of beam shape and for this reason all software proximity effect corrections were disabled during this experiment. Abundant research has been put into the measurement of the beam size and shape using the point spread function (PSF) method.[20] Here, we adopt the widely used notation and define the beam shape as the sum of two Gaussians, the first representing the highly collimated primary beam, and another representing the contribution from backscattered electrons:

$$I_{EBL}(x) = \frac{1}{2\alpha^2}e^{-x^2/\alpha^2} + \frac{\eta}{2\beta^2}e^{-x^2/\beta^2} \quad (2)$$

where α is the square root of the variance of the forward scattering beam, β is the square root of the variance of the backscattering electrons and η is defined as a correction factor. Experimentally measured values for α range from 4 to 14 nm, whereas β is about several microns; the correction factor η is typically ≈ 0.7-1.[21,22,23] Based on literature data, the amplitude of the primary Gaussian ($1/2\alpha^2$) can be estimated to be about 100 to 1000 times larger than the amplitude of the secondary Gaussian ($\eta/2\beta^2$). The expression of Eq. (2) therefore describes the combination of a sharp beam profile with a very broad and low-intensity tail.

## C.   He ion beam lithography

A He ion beam was generated by field ionization from a gas source in a ZEISS microscope column, and accelerated to 30 keV. The raster scan was controlled by an Orion Nanofab pattern generator to perform direct write lithography of arbitrary user-defined layout. The beam aperture was 7 μm and the write current was 0.19 pA for all samples. Although the source and the column are capable of supplying higher currents, these settings were dictated by the need for high resolution and by the maximum allowed frequency of the beam blanking unit. In this equipment, the raster scan speed is possibly the main disadvantage of the direct write with focused beam, as less than a square micron per second can be patterned, using these settings for high resolution.

Early works demonstrated that the large mass and large scattering cross section of helium ions lead to a much shallower penetration depth in matter and a smaller interaction volume than it occurs with electron beam, thus bringing remarkably superior imaging performance.[24,25] Similar to the electron beam, the PSF of the He ion beam has been determined by previous investigations and it has been conventionally modeled as the sum of two Gaussians.[26,27] As these studies found, the experimental measurement of the PSF was significantly more challenging than for electron beam, because the He beam size is well below the resolution limit of printing detection and its proximity effect is weaker. Among the few works on this topic, one estimated the PSF indirectly by Abel inversion of experimentally measured line spread function of a chemically amplified resist.[28] Reported r.m.s. width values are 0.9 nm and 150 nm for the primary and secondary beams, respectively.[29] Notably, the intensity of the secondary beam was estimated to be six orders of magnitude weaker than that of the primary beam. For this reason, in this work we modeled the aerial image of the ion beam as a single Gaussian:

$$I_{HIBL}(x) = \frac{1}{2\sigma^2} e^{-x^2/\sigma^2} \qquad (3)$$

where σ is the Gaussian standard deviation.

## D.   Resist materials

Poly-methylmethacrylate (PMMA, molecular weight 950k, 1% w/w in ethyl lactate) and hydrogen silsesquioxane (HSQ, 1% in methyl isobutyl ketone, MIBK) were chosen because they are sensitive resists over a broad range of energies and patternable by both photons and particle beams. For the sake of comparison, the processing was kept the same throughout all lithographic tools: thermal treatments, development time, and development conditions. Both resists were spin coated to target a 40 nm film thickness. A post application bake of 180 °C and 5 minutes was used for PMMA, while no thermal treatment was used for HSQ. PMMA was developed in methyl isobutyl ketone and isopropyl alcohol 1:3 mixture (MIBK:IPA) for 50 seconds and rinsed for 30 s in pure IPA. HSQ was developed in tetramethylammonium hydroxide and deionized water 1:3 mixture (TMAH:DIW) for 30 s, followed by rinsing in deionized water for 30 s, and isopropyl alcohol for 30 s. All developments were done at room temperature.

## III. MODELING

The Lumped Parameters Model (LPM)[15] is based on the segmented development of a resist after exposure and it describes the geometry of the resist image resulting from a given aerial image. The LPM belongs to the category of 'simplified resist models', because it lumps (as the name implies) the effects of several physical mechanisms into a limited number of parameters. The LPM is also an analytical model: it is based on the combination of the resist development kinetics with the exposure intensity distribution on and in the resist. As a consequence, this model is very compact and has a low computational burden. In the specific case of two dimensional l/s symmetry, its exact analytical form is given by the expression:[30]

$$E(x) = E_0 \left[ 1 + \frac{1}{D_{eff}} \int_{x_0}^{x} \left( \frac{I(x')}{I(x_0)} \right)^{-\gamma} dx' \right]^{\frac{1}{\gamma}} \quad (4)$$

where $E$ is the energy (or dose) required to print a line with CD of $x$, by an aerial image of normalized intensity $I(x)$ on a resist of contrast $\gamma$ and dose-to-clear $E_0$; $x_0$ indicates the center location of the line. Because this model was devised specifically for photolithography, an effective thickness $D_{eff}$ in Eq. (4) is defined as:

$$D_{eff} = \frac{1}{\alpha\gamma}(1 - e^{-\alpha\gamma D}) \qquad (5)$$

and it accounts for the finite transmissivity of the material via the actual resist thickness *D*, its optical absorption coefficient *α*, and resist contrast *γ*.

The LPM is, therefore, a compact and accurate method to quantitatively calculate the CD vs. Dose relationship. An important aspect of the LPM is that it takes into account and can be tuned for different aerial images. A MATLAB algorithm of the exact LPM model of Eq. (4) was implemented using the numerical integration method (global adaptive quadrature). In a conventional experiment, the CD of lines is measured after an exposure of known dose. To use the LPM, CD values ranging from about 0 to pitch are inputted in Eq. (4) and the corresponding dose is extracted. The aerial images *I(x)* are then replaced with the definitions of Section II for the three lithographic tools. Because different tools have specific dose-to-clear, for the sake of this comparison, the CD vs. dose plots was normalized to $E_0$. Aside from the obvious differences in exposing a material using a photon beam, an electron beam and an ion beam, the exposure reaction is always triggered by electrons. In the case of EUV-IL, these electrons are secondary electrons (SE) generated by ionization following the optical absorption. In the case of EBL and HIBL, both the primary beam and the SE generated by the impact ionization of the primary do contribute to the reaction. As a result, there is a difference in the spatial energy distribution of generated SE, as it was demonstrated by a comparative study of chemically amplified resists exposed by EBL and EUV. Those Authors found a that in the former, ionization occurs mostly in spatially isolated reactions, whereas in the latter, the probability of multiple ionization events in confined space is about a factor three-fold higher.[31] The LPM model does not account for the spatial distribution of ionization events because only the aerial image intensity (i.e., its spatial energy) is provided as an input.

The LPM was employed to calculate the CD vs. dose for three aerial image conditions: sinusoidal beam (EUV-IL), Gaussian beam with proximity effect (EBL), and single Gaussian beam with no proximity effect (HIBL) when patterning dense l/s on an arbitrary positive tone resist of contrast $\gamma = 1$, absorption coefficient $\alpha = 5$ μm$^{-1}$ and thickness 20 nm. In the outlook of cutting edge processing, we considered the hypothesis of patterning dense l/s patterns of pitch 20 nm, which presents some challenge for current state-of-the-art lithography (analogous results hold valid for less stringent resolution). Based on the aforementioned studies of the PSF of EBL and HIBL, the width of the primary beam (σ) was assumed to be 5 nm. To simplify the expression of the $I_{EBL}(x)$ aerial image, the proximity effect of the EBL which arises from a dense array of lines is cumulatively incorporated into a single term of the secondary Gaussian. Therefore, the r.m.s. amplitude of the secondary Gaussian was set to 1/50 of the amplitude of the primary beam and its width equal to σ = 5 μm. With this setting, the aerial image of the EBL simplifies and approximatively accounts for the cumulative proximity arising from 100 adjacent lines having a secondary beam intensity equal to 1/5000 of the primary. A comparison of the calculated CD for the three lithographic tools is then shown in the following Fig. 1, where the dose was normalized to each tool's specific dose-to-clear $E_0$, and the CD was normalized to the pitch.

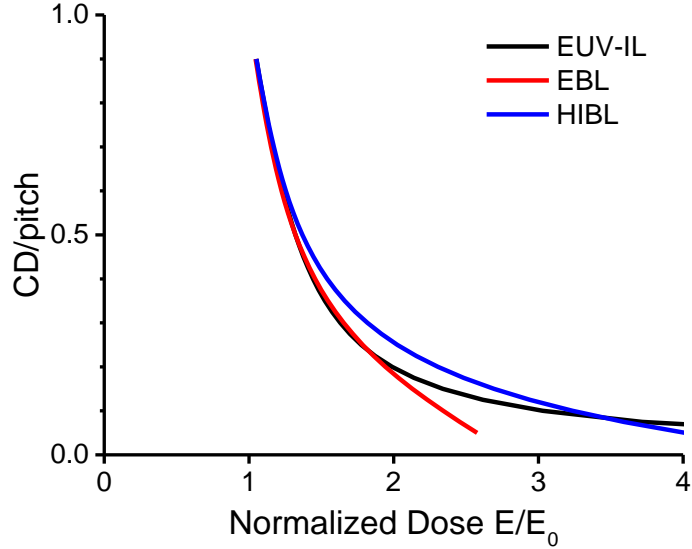

FIG. 1. (Color online) Critical dimension, normalized to the pitch, as a function of normalized dose calculated by using the LPM for dense l/s of pitch 20 nm exposed by EUV-IL (black line), EBL (red line) and HIBL (blue line). The parameters for the LPM were set to a positive tone resist of thickness 20 nm, contrast 1 and absorption coefficient 5 $\mu m^{-1}$. The $E_0$ dose represents the resist's dose-to-clear for each specific tool.

The different trends of the CD vs. dose in Fig. 1 represent exclusively the effect of the aerial image, while all the other parameters are the same. The LPM predicts, as expected, that no lines are printed at dose below the $E_0$ threshold. As the dose increases above $E_0$, the estimated CD decreases (as a consequence of the increase of the width of the cleared trench, being it a positive tone resist) in all cases, although with different slope. The CD of lines patterned using EBL and HIBL decreases until the lines become overexposed and the CD gets to zero. However, this saturation point is reached much faster in the former case than in the latter, owing to the contribution from proximity effect which effectively contributes to the broadening of linewidth. In the case of ion beam (and in general any purely Gaussian aerial intensity) the CD varies weakly with the dose, and the lines merge together only at very high energies, when the effect of the tail of the Gaussian proximity effect from adjacent lines increases in magnitude. Due to the lack of proximity effect, the ion beam also requires a relatively higher dose to pattern lines of the same CD as those patterned by EBL. Finally, the CD of lines printed using EUV-IL levels off at higher doses and never completely saturate, even at higher energy. This observation is a direct consequence of the aerial image intensity $I_{EUV-IL}(x)$ being exactly zero at the center of the line geometry, regardless of the dose and pitch. It is important to clarify that the LPM estimations are as accurate as the respective PSF, on which they are based. This observation becomes especially relevant at relatively high doses, when the determination of the tail of the PSF is inaccurate owing to the solubility change of the resist.

## IV. RESULTS AND DISCUSSION

### A. *Resist Dose-to-Clear ($E_0$) and Contrast (γ) parameters*

The $E_0$ of PMMA (i.e. the dose-to-clear) and of HSQ (i.e. the dose-to-gel) were measured by exposing arrays of large squares at increasing dose. (In the lithographic community, the dose-to-clear is usually indicated by the symbol $D_0$: in this work, $E_0$ will be used for consistency with the LPM notation.) The measurement of $E_0$ provides information on the relative efficiency of a lithographic equipment in patterning a given material. In addition, the $E_0$ is also useful as parameter for the LPM fit as mentioned above. In the case of EUV-IL, 0.5×0.5 mm$^2$ open frame exposures were conducted and the remaining thickness was measured by profilometer. In a similar way, large areas of size in excess of 250 μm$^2$ areas were patterned in the case of EBL and HIBL; the remaining thickness was measured by atomic force microscopy in contact mode. The contrast curves, normalized to the film thickness, are shown in Fig. 2. The experimental data were then fitted to a dose response curve, from which the $E_0$ and γ were extracted; these values are summarized in Table I.

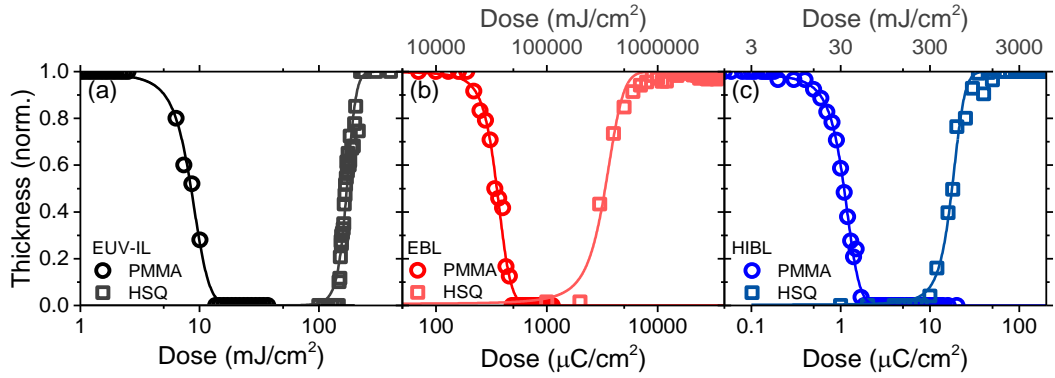

FIG. 2. (Color online) Experimentally measured contrast curves for PMMA (circles) and HSQ (squares) exposed by EUV-IL (a), EBL (b), and HIBL (c). The solid lines are dose response fits to the corresponding data and were used to extract the dose-to-clear $E_0$ and contrast $\gamma$ of the resists for each specific exposure conditions. For EBL and HIBL, the equivalent dose in unit energy per area is also reported on the top axis.

In agreement with the known properties of these materials, our data indicate that HSQ is significantly less sensitive than PMMA: the $E_0$ was more than one order of magnitude higher, regardless of the tool. As for the 'relative sensitivity' of these tools when exposing the same resist, a considerable gap can be detected in the amount of charge per unit area required to clear. The EBL required almost a two orders of magnitude higher dose than the HIBL did. Several studies had previously reported a similar observation: the dose-to-clear obtained by the former is about 4-fold to hundred-fold higher than the latter.[5,8,9,32] The exposure efficiency is described by the secondary electron yield (SEY) parameter, which indicates the amount of secondary electrons generated by each absorbed primary photon, electron or ion. This definition is widely used in the photoresist community to describe the efficiency of each SE generation: it is not related to the total incident dose nor to the surface effects, but to the dose absorbed throughout the resist. According to our experimental data, the dose required to fully expose these materials by the HIBL is a 100-fold lower than that needed by the EBL, for both resists. Other authors estimated the SEY of He ion beam to be between 7 and 30.[33,34] Those estimations are in agreement with the experimental observations of a decreasing sensitivity of these materials when using a 100 keV electron beam, as in the present work, as compared to the 30 keV and lower energies reported in previous works.[9]

Furthermore, the energetic efficiency of the EUV-IL can be discussed by calculating the equivalent dose for EBL and HIBL in unit energy per area, based on the kinetic energy of these tools and shown in the dosage curves of Fig. 2, top axis. It can be noticed that the ion beam required only about 3 times more energy than the EUV photons to clear PMMA, whereas the EBL needed an incident dose as much as 3000-fold higher. Furthermore, the dose gap between HIBL and EBL is, in terms of incident energy, wider than it was in terms of deposited charge. In other words, the relative energetic efficiency of EBL is lower than the simple charge dosage would suggest. This finding is consistent with the well-known fact that the EBL dose is inversely proportional to the acceleration voltage. In summary, the electron beam exhibits a remarkably low lithographic efficiency as consequence of both the weaker interaction with matter and the high kinetic energy (100 keV) as compared to that of the ion beam (30 keV).

While many studies have previously experimentally compared EBL with HIBL, there is a remarkable lack of systematic studies of electron beam exposures versus extreme ultraviolet lithography. Recent works by Kyser et al.[35] and by Oyama et al.[36] both proposed two approaches to calibrating the dose ratio between EBL and EUV exposures, expressed in $(\mu C/cm^2)/(mJ/cm^2)$. From the calculated deposited dose per unit volume in the resist, this dose ratio was estimated to be $\approx 7$ (in the case of 100 keV EBL)[35] and ≈ 3.5 (in the case of a 75 keV EBL).[36] However, dose ratios reported in experimental studies vary from 4.5,[37] ≈ 10,[38] and between 7 and 35.[39] This ratio, calculated from our dose-to-clear of PMMA and HSQ, is ≈ 42 and ≈ 20, respectively. It is not straightforward to explain the discrepancy between theory and experiment, and among all these experimental values as well. It should be noted that those theoretical estimates assume that the dose required to fully expose a resist depend exclusively on the energy deposited (based on the equivalency of the volumetric absorbed dose, in the case of EUV, and of the primary e-beam electron in the case of EBL). However, previous works have shown that the ionization efficiency might differ, as it was demonstrated in the case of the volumetric probability of photoacid generation. Exposure by EUV, for instance, is more likely to produce multiple ionization event in a confined volume of resist than the EBL can accomplish.[40] The experimental dose ratio reported in the present work is higher than the theoretical predictions, which we ascribe to a higher lithographic efficiency of EUV than it would be expected from the pure dose equivalency. Finally, the broad differences in the dose ratio across different resist platforms reported in one of the previous works,[39] is suggesting that the acid generation probability is also a specific variable in the resist exposure kinetics. While most of the above mentioned values are extracted from l/s patterning at the 1:1 duty cycle dose (i.e., the dose-to-size), our dose ratio are instead calculated from dosage curves: this difference does not changes our argumentation.

As for the resist contrast, this quantity is strongly dependent on the processing conditions and previous studies reported values of $\gamma$ ranging from 1 to 3.5 for both PMMA and HSQ materials developed using conventional methodology.[9,27,41] According to our data, the contrast was not changed significantly across the lithographic tools, which is a good indication of the consistency of the processing, despite being produced by different equipment.

TABLE I. Summary of measured dose-to-clear $E_0$ and contrast γ for all resists and tools investigated in this work.

|  | Dose-to-Clear $E_0$ | | | Contrast γ | | |
| --- | --- | --- | --- | --- | --- | --- |
|  | EUV-IL [mJ/cm$^2$] | EBL [μC/cm$^2$] | HIBL [μC/cm$^2$] | EUV-IL | EBL | HIBL |
| PMMA | 8.5 | 355 | 1.1 | 1.4 | 1.6 | 1.5 |
| HSQ | 171 | 3364 | 17.6 | 1.8 | 1.3 | 1.2 |

## *B. Dense lines/spaces*

The performance in patterning dense l/s was tested using EUV-IL, EBL and HIBL on PMMA (pitch 80 and pitch 60 nm) and HSQ (pitch 44 nm). The CD of lines was measured by metrological scanning electron microscopy. The resulting CD, normalized to the total pitch, is plotted for these tools as a function of the normalized dose $E/E_0$ in Fig. 3.

A consistent trend can be observed from these plots. At normalized dose below 1, that is, when $E < E_0$, no lines are printed: $E_0$ represents a minimum threshold energy. As the dose is increased, in a positive tone resist as it is PMMA, trenches begin to be printed and the measured CD decreases with increasing dose: the measured CD represents here the width of remaining resist lines between the trenches (measured by the SEM). As for HSQ, the resulting CD indicates a similar behavior as PMMA. In the case of this negative tone resist, the CD of lines increases with increasing dose.

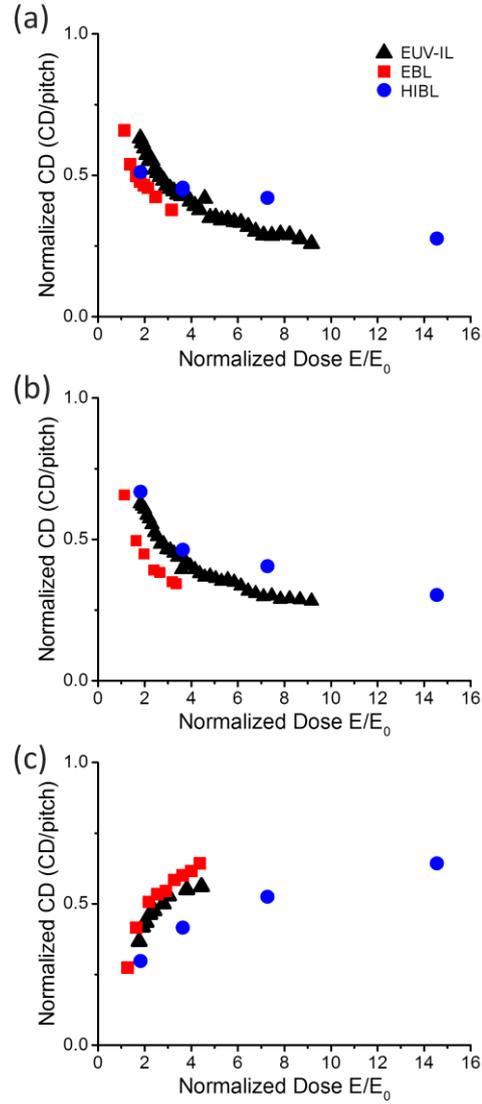

FIG. 3. (Color online) Normalized critical dimension as a function of normalized dose for dense l/s, 80 nm pitch (a) and 60 nm pitch (b) in PMMA, and of 44 nm pitch in HSQ (c), exposed by EUV-IL (black triangles), EBL (red squares), and HIBL (blue circles).

Remarkable differences in the CD of lines printed by these tools can be noticed. At higher doses, the CD patterned by EUV-IL levels off and never saturates. For EBL, the CD increases faster with dose than it does in other tools; while the ion beam exposes lines have an almost linear trend with dose. These same trends occur regardless of the pitch and of the resist tone. Moreover, the onset of the threshold depends on the tool used. Considerations on the different features of the tools investigated here will follow to explain the different graph behavior now. In the EUV-IL, the dose intensity is formed by an interference pattern and it is ideally a sinusoidal curve: therefore, a zero intensity point is always present, regardless of the amplitude (i.e. dose). As a result, in a dense l/s pattern, adjacent lines will ideally always be well detached from each other (neglecting here the effect of resist blur). This was experimentally confirmed by the leveling off of the EUV-IL trace. On the opposite, patterning dense l/s by EBL and HIBL produces a different result due to the proximity effect of the beam in adjacent lines. In the case of the EBL, this effect is much more marked due to the broad spatial effect of the backscattered electrons. This observation was confirmed by the steeper decrease or increase of the EBL traces compared to the $He^+$ traces for PMMA and HSQ, respectively. In conclusion, it can be qualitatively understood that it is the beam geometry that affects the spatial energy distribution and thus the CD vs. dose behavior.

## C. LPM fit to EUV-IL

The lumped parameter model provides a compact and precise estimation of the CD vs. dose relationship which is of interest in the scope of this work. In this section we aim at validating our MATLAB implementation of the LPM against experimental data. EUV-IL exposures of dense l/s were carried out on PMMA (pitches 100, 80 60 and 44 nm) and on HSQ (pitch 44, 36, 32 nm). Because the EUV-IL mask used to this purpose features all of the above mentioned periodicities, all the l/s pitches were exposed at once on the same wafer. Subsequently, the entire sample was processes at the same time and under the same conditions for all pitches, thus providing a consistent comparison. The exposure dose was varied so as to obtain as many data point as possible in a broad range.

A nonlinear least-squares fit routine, based on the Levenberg-Marquardt algorithm, was implemented in MATLAB for the regression of the LPM expression of Eq. (4) to the experimental CD vs. dose data. To this purpose, the normalized aerial image was set to the function $I_{EUV-IL}(x)$ described in Eq. (1), Section II, and the period $p$ was set to the corresponding data. The absorption coefficient $\alpha$ was set to 5.0 μm$^{-1}$, based on previous measurement of the optical absorption of PMMA and HSQ at EUV wavelength.[42] The dose were normalized to the $E_0$ of the two materials at EUV, as measured in Section IVa. The regression algorithm provided the $\gamma$ which best fitted to the data. The resulting best fit LPM are shown alongside the experimental CD data in the following Fig. 4, and the values are summarized in Table II.

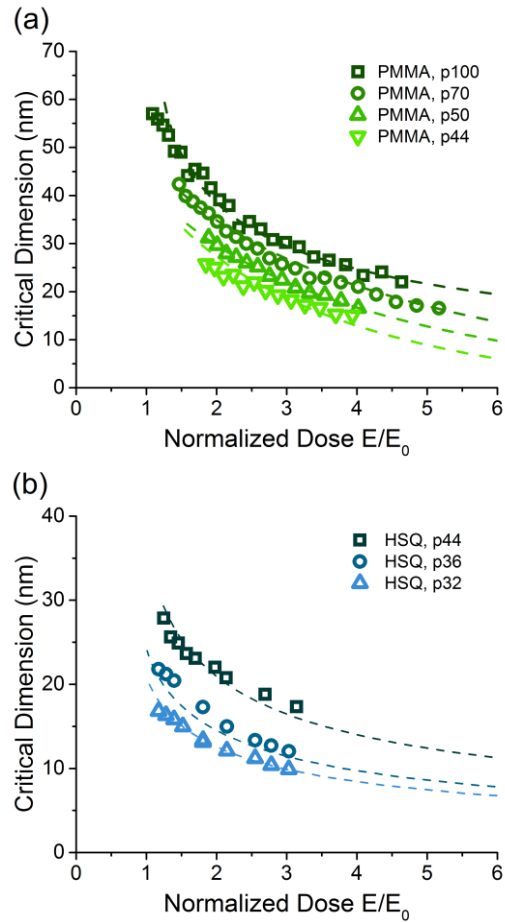

FIG. 4. (Color online) Experimental critical dimension (symbols) of dense lines/spaces as a function of normalized dose, patterned by EUV-IL on PMMA (a) and HSQ (b). e The corresponding nonlinear least-squares fit to the each data set, obtained using the Lumped Parameter Model of Eq. (4) is also shown (dashed lines).

The regression algorithm provided a good fit throughout all the range. In the proximity of the dose-to-clear ($E/E_0 \approx 1$), the fit diverges from the data, which can be ascribed to the inaccuracy in measuring the critical dimension of very underexposed lines. It was found that the extracted contrast $\gamma$ was below the value measured from contrast curves. In this regard, the contrast of a resist is influenced by the development contrast $\gamma_d$ (which depends on the processing) and by the exposure contrast $\gamma_e$ (which depends on the beam decay in the resist due to optical absorption), according to the relation $\gamma_t^{-1} = \gamma_d^{-1} + \gamma_e^{-1}$.[30] In our implementation of the LPM, we employed the effective thickness $D_{\mathit{eff}}$, as mentioned in Section III. As a result, the $\gamma$ parameter extracted from the LPM fit is already accounting for the detrimental effect of both the absorption and the non-ideal development. This observation seems to indicate a loss of lithographic contrast of the resist when patterned in dense l/s, in comparison to the same resist patterned in open frame exposures. These two non-idealities concurrently decrease the contrast of the high resolution patterns. Finally, it must be noticed that the contrast values did not show any dependence on the pitch of the l/s: as can be seen from the slope of the fitting, which is similar throughout all pitches. We ascribe this effect to the quality of the aerial image of EUV-IL, which is described by a constant normalized image-log slope value of $\pi$, regardless of the pitch.[43]

TABLE II. Summary of best fit parameters using the LPM regression to the EUV-IL exposures of PMMA and HSQ dense l/s.

| pitch | γ (PMMA) | γ (HSQ) |
|---|---|---|
| 100 nm | 0.45 | - |
| 80 nm | 0.50 | - |
| 60 nm | 0.41 | - |
| 44 nm | 0.46 | 0.65 |
| 36 nm | - | 0.60 |
| 32 nm | - | 0.62 |

## V. SUMMARY AND CONCLUSIONS

A comparative study of the extreme ultraviolet interference lithography, electron beam lithography and He ion beam lithography has been presented here. Despite the different working principle, the similarity in the exposure mechanism in broadband resists made it possible to conduct this comparison. Preliminarily, we determined the dose-to-clear and the resist contrast of PMMA and HSQ for each tool, which was used to normalize the exposure doses. In agreement with previous studies, it was found that the He ion beam lithography required about a hundred-fold lower dose than electron beam lithography did. From the dosage point of view, our findings indicate that EBL is more than four orders of magnitude less lithographically efficient than EUV-IL and more than three orders of magnitude less than HIBL, owing to the weak interaction with the resist and the high kinetic energy of the electron beam. The ratio between the exposure doses by EBL and by EUV-IL, found experimentally in the present work, was higher than theoretical estimates reported in previous studies.

We discussed this discrepancy in the light of the different in ionization efficiency of these lithographic tools.

The critical dimension of dense l/s patterns, which is a relevant figure of merit for lithographic processing and fabrication of integrated circuits in general, was studied in detail. The beam shape had a remarkable effect on the CD vs. dose relationship, as it accounts for the specific features of each tool. A numerical formulation of the exact lumped parameter model and a nonlinear least-squares regression were also implemented. This model estimated quantitatively the effect of different aerial images on the CD vs. dose, consistently with the experimental findings. Finally, the LPM fit to the EUV-IL exposure data extracted the parameters for $\gamma$ and $E_0$, thus validating our implementation. Our findings quantitatively explain for the peculiar features each tool has and which make it suitable for different purposes. The EUV-IL make it possible to pattern large areas of dense features such as line/spaces, with relatively good resolution. The electron beam lithography is effective in exposing high resolution arbitrary patterns and has a better performance for isolated structures than dense ones. The HIBL is a promising technique for the lithography of dense high resolution patterns due to the almost negligible backscattered secondary electrons from the beam-substrate interaction, which possibly make it ideal for both isolated and dense high resolution patterning.

## ACKNOWLEDGMENTS

We gratefully acknowledge Simon Tschupp (PSI) for sample analysis and fruitful discussion and Vitaliy Guzenko (PSI) and Michaela Vockenhuber (PSI) for fruitful discussion. This project has received funding from the EU-H2020 Research and Innovation programme under grant agreement No. 654360 NFFA-Europe. One of the Authors (R.F.) acknowledges Inpria Corp. for funding.

[1] D. R. Herriott, J. Vac. Sci. Technol. **20**, 781 (1982).

[2] K. Wilder, C. F. Quate, B. Singh, and D. F. Kyser, J. Vac. Sci. Technol. B **16**, 3864 (1998).

[3] S. Y. Chou, P. R. Krauss, and P. J. Renstrom, Appl. Phys. Lett. **67**, 3114 (1995).

[4] W. L. Brown, T. Venkatesan, and A. Wagner, Nucl. Instrum. Methods **191**(1), 157 (1981).

[5] S. Hirscher, R. Kaesmaier, W.-D. Domke, A. Wolter, H. Loeschner, E. Cekan, C. Horner, M. Zeininger, and J. Ochsenhirt, Microelectron. Eng. **57**, 517 (2001).

[6] M. Chan, R. R. Kunz, S. P. Doran, and M. Rothschild, J. Vac. Sci. Technol. B **15**, 2404 (1997).

[7] V. Bakshi, EUV lithography, 1st ed. (SPIE Press, Bellingham, Washington USA, 2009), pp.42-43.

[8] J. Notte, B. Ward, N. Economou, R. Hill, R. Percival, L. Farkas, and S. McVey, AIP Conf. Proc. **931**, 489 (2007).

[9] V. Sidorkin, E. van Veldhoven, E. van der Drift, P. Alkemade, H. Salemink, and D. Maas, J. Vac. Sci. Technol. B **2**(4), 18 (2009).

[10] C. A. Sanford, L. Stern, L. Barriss, L. Farkas, M. DiManna, R. Mello, D. J. Maas, and P. F. A. Alkemade, J. Vac. Sci. Technol. B **27**(6), 2660 (2009).

[11] D. Winston, V. R. Manfrinato, S. M. Nicaise, L. L. Cheong, H. Duan, D. Ferranti, J. Marshman, S. McVey, L. Stern, J. Notte, and K. K. Berggren, Nano Lett. **11**, 4343 (2011).

[12] F. H. M. Rahman, S. McVey, L. Farkas, J. A. Notte, S. Tan, and R. H. Livengood, Scanning **34**, 129 (2012).

[13] E. A. Dobisz, H. W. P. Koops, F. K. Perkins, C. R. K. Marrian, and S. L. Brandow, J. Vac. Sci. & Technol., B **14**, 4148 (1996).

[14] M. P. C. Watts, J. Vac. Sci. & Technol., B **3**, 434 (1985).

[15] N. G. Einspruch, Lithography for VLSI: VLSI Electronics - Microstructure Science (Academic Press, New York, USA, 1987), pp. 19–55.

[16] Z. Liu, F. Bouamrane, M. Roulliay, R. K. Kupka, A. Labèque, and S. Megtert, J. Micromech. Microeng. **8**(4), 293 (1998).

[17] A. DeHon, IEEE T. Nanotechnol. **2**(1), 23 (2003).

[18] E. Buitrago, R. Fallica, D. Fan, T. S. Kulmala, M. Vockenhuber, and Y. Ekinci, Microelectron. Eng. **155**, 44 (2016).

[19] D. Fan, and Y. Ekinci, J. Micro-nanolith. Mem. **15**(3), 033505 (2016).

[20] T. H. P. Chang, J. Vac. Sci. & Technol. **12**, 1271 (1975).

[21] S. A. Rishton and D. P. Kern, J. Vac. Sci. & Technol., B **5**, 135 (1987).

[22] M. Osawa, K. Takahashi, M. Sato, Masami and H. Arimoto, K. Ogino, H. Hoshino, and Y. Machida, J. Vac. Sci. & Technol., B **19**, 2483 (2001).

[23] H. Duan, V. R. Manfrinato, J. K. W. Yang, D. Winston, B. M. Cord, and K. K. Berggren, J. Vac. Sci. & Technol., B **28**(6), C6H11 (2010).

[24] D. C. Joy and S. Luo, Scanning **11**(4), 176 (1989).

[25] D. Cohen-Tanugi and N. Yao, J. Appl. Phys. **104**, 063504 (2008).

[26] D. Winston, B. M. Cord, B. Ming, D. C. Bell, W. F. DiNatale, L. A. Stern, A. E. Vladar, M. T. Postek, M. K. Mondol, J. K. W. Yang, and K. K. Berggren, J. Vac. Sci. & Technol., B **27**, 2702 (2009).

[27] G. Hlawacek, V. Veligura, R. van Gastel, and B. Poelsema, J. Vac. Sci. & Technol., B **32**, 020801 (2014).

[28] N. Kalhor, W. Mulckhuyse, P. Alkemade, and D. Maas, Proc. SPIE Conf. **9425**, 942513 (2015).

[29] D. Winston, J. Ferrera, L. Battistella, A. E. Vladár, and K. K. Berggren, Scanning **34**(2), 121 (2012).

[30] C. Mack, Fundamental Principles of Optical Lithography (John Wiley & Sons, Chichester, West Sussex, England, 2007), p. 286.

[31] T. Kozawa, K. Okamoto, A. Saeki, and S. Tagawa, Jap. J. Appl. Phys. **48**, 056508 (2009).

[32] L. Scipioni, and D. Winston, Helium Ion Beam Lithography in the ORION® PLUS, Zeiss Whitepaper, July 2009.

[33] J. Morgan, J. Notte, P. Hill, and B. Ward, Micros. Today **14**, 24 (2006).

[34] D. Maas, E. van Veldhoven, A. van Langen-Suurling, P. F.A. Alkemade, S. Wuister, R. Hoefnagels, C. Verspaget, J. Meessen, and T. Fliervoet, Proc. SPIE Conf. **9048**, 90482Z (2014).

[35] D. F. Kyser, N. K. Eib, and N. W. M. Ritchie, J. Micro-nanolith. Mem. **15**(3), 033507 (2016).

[36] T. G. Oyama, A. Oshima, and S. Tagawa, AIP Adv. **6**, 085210 (2016).

[37] S. Bhattarai, A. R. Neureuther, and P. P. Naulleau, Proc. SPIE Conf. **9422**, 942209 (2015).

[38] J. Stowers, A. Telecky, M. Kocsis, B. L. Clark, D. A. Keszler, A. Grenville, C. N. Anderson, and P. P. Naulleau, Proc. SPIE Conf. **7969**, 796915-1 (2011).

[39] L. D. Bozano, P. J. Brock, H. D. Truong, M. I. Sanchez, G. M. Wallraff, W. D. Hinsberg, R. D. Allen, M. Fujiwara, and K. Maeda, Proc. SPIE Conf. **7972**, 797218-1 (2011).

[40] T. Kozawa and S. Tagawa, Jap. J. Appl. Phys. **49**, (2010) 030001.

[41] J. K. W. Yang, B. Cord, H. Duan, K. K. Berggren, J. Klingfus, S.-W. Nam, K.-B. Kim, and M. J. Rooks, J. Vac. Sci. & Technol., B **27**, 2622 (2009).

[42] R. Fallica, J. K. Stowers, A. Grenville, A. Frommhold, A. P. G. Robinson, and Y. Ekinci, J. Micro-nanolith. Mem. **15**(3), 033506 (2016).

[43] A. Langner, H. H. Solak, R. Gronheid, E. van Setten, V. Auzelyte, Yasin Ekinci, K. van Ingen Schenaud, and K. Feenstrad, Proc. SPIE Conf. **7636**, 76362X-1 (2010).